%% file: 0_main.tex
\documentclass[5p]{elsarticle}

\usepackage{lineno,hyperref}
\usepackage{color} % for draft  %tomo
\usepackage{amsmath}
\journal{Nuclear Inst. and Methods in Physics Research, A}
\usepackage[hang,small,bf]{caption}
\usepackage[subrefformat=parens]{subcaption}
\begin{document}

\begin{frontmatter}

\title{Comparative pulse shape discrimination study for Ca(Br, I)$_2$ scintillators using machine learning and conventional methods}

\author[address2]{M. Yoshino\corref{mycorrespondingauthor}}
\cortext[mycorrespondingauthor]{Corresponding author}
\ead{yoshino.masao@tohoku.ac.jp}

\author[address1]{T.~Iida}
\author[address4]{K.~Mizukoshi}
\author[address5]{T.~Miyazaki}
\author[address3,address6]{K.~Kamada}
\author[address2]{K.~J.~Kim}
\author[address2,address3,address6]{A.~Yoshikawa}

\address[address2]{Institute for Material Research, Tohoku University, Miyagi 980-8577, Japan}
\address[address1]{Faculty of Pure and Applied Sciences, University of Tsukuba,  Tsukuba, Ibaraki, 305-8571, Japan}
\address[address4]{Department of Physics, Kobe University, Kobe, Hyogo 657-8501, Japan}
\address[address5]{Graduate School of Engineering, Tohoku University, Sendai, Miyagi 980–8579, Japan}
\address[address3]{New Industry Creation Hatchery Center, Tohoku University, Sendai, Miyagi 980-8579, Japan}

\address[address6]{C\&A Corporation, Sendai, Miyagi 980-0811, Japan}

%\author{Masao Yoshino\fnref{myfootnote}}
%\address{Sendai}
%\fntext[myfootnote]{aaa}




\begin{abstract}
In particle physics experiments, pulse shape discrimination (PSD) is a powerful tool for eliminating the major background from signals. However, the analysis methods have been a bottleneck to improving PSD performance. 
In this study, two machine learning methods---multilayer perceptron %network
and convolutional neural network---were applied to PSD, and their PSD performance was compared with that of conventional analysis methods.
Three calcium-based halide scintillators were grown using the vertical Bridgman--Stockbarger method and used for the evaluation of PSD. Compared with conventional analysis methods, the machine learning methods achieved better PSD performance for all the scintillators.
For scintillators with low light output, the machine learning methods were more effective for PSD accuracy than the conventional methods in the low-energy region.
\end{abstract}


\end{frontmatter}

%\linenumbers

\input{1_intro}
\input{2_1_material}
\input{2_2_method}
\input{2_3_ml}

\input{3_result}
\input{5_conclusion}

\section*{Acknowledgement}
%This work was supported by JSPS KAKENHI Grant-in-Aid for Scientific Research (B) 18H01222, 21H03834 and Grant-in-Aid for Young Scientists (B), Japan 16K17700. 
This work was supported by JSPS KAKENHI [grant numbers 18H01222, 21H03834, 16K17700, 19K11848]. %tomo
It was performed as part of the Inter-University Cooperative Research Program of the Institute for Materials Research, Tohoku University [proposal numbers 19K0089, 20K0035, 202012-RDKGE-0016]. 
This work was partially supported by the KEK Detector R\&D Platform. We thank Maxine Garcia, PhD, from Edanz (https://jp.edanz.com/ac) for editing a draft of this manuscript.

% \section*{References}

\bibliography{mybibfile}

\end{document}

%% file: 1_intro.tex
\section{Introduction} \label{sec:intro}
Radiation detectors that use inorganic scintillators have been widely used in basic science research, as well as in applied fields such as medicine and industry, because of their high energy resolution and high density.
The discrimination of particle types, such as $\gamma$-rays, $\alpha$-rays, and neutrons, from the obtained signals is a fundamental technique in particle experiments that require a low background environment, including the search for dark matter and neutrino-less double beta decay.
However, it is difficult to achieve high discrimination power using conventional analysis techniques that use differences in the shape of the rising edge and tail of the waveform for each particle.
To increase sensitivity, it is necessary to establish a superior analysis technique using crystals that have a large amount of scintillation light yield and a large difference in waveform between particle types.

Detectors with excellent particle discrimination include gas-liquid dual-phase detectors \cite{gas}, which use the noble gas xenon to measure two types of signals, that is, the ionization signal and scintillation light; and scintillating bolometer detectors \cite{bolometer}, which use inorganic crystals chilled to cryogenic temperatures to measure two types of signals, that is, the thermal signal and scintillation light.
These detectors are not suitable for general applications in medicine and industry because they require expensive and large equipment.

We conducted a comparative study of particle identification by waveform using our own calcium-based halide scintillators (CaI$_2$, CaBr$_{0.7}$I$_{1.3}$, and Eu2\% doped CaBr$_{0.7}$I$_{1.3}$).
In our previous research, CaI$_2$ achieved a high light yield and excellent energy resolution \cite{CaI2}. Additionally, we reported the good pulse shape discrimination (PSD) capability of CaI$_2$ using a simple double gate method \cite{CaI-PSD}.
CaBr does not have a cleavage plane, but does have low light output; therefore, we developed and reported the growth and scintillation properties of Eu doped- and undoped-CaBrI crystals \cite{CaBrI}.
Because CaI$_2$ contains $^{48}$Ca, which is a double beta decay nucleus, we plan to apply this crystal to the search for the double beta decay of Ca in the future.

In this study, we apply machine learning-based analysis methods to achieve particle identification with high sensitivity. The PSD methods evaluated in this study are the double gate (DG) method, digital filter (DF) method, and two machine learning methods: the multilayer perceptron (MLP) %network 
and convolutional neural network (CNN).
In this paper, we discuss the PSD performance of three scintillators using the four aforementioned analysis methods.

%% file: 2_1_material.tex
\section{Materials and Methods} \label{sec:mat}
\begin{figure*}[ht]
    \begin{tabular}{ccc}
      \begin{minipage}[t]{0.3\linewidth}
        \centering
        \includegraphics[keepaspectratio, scale=0.3]{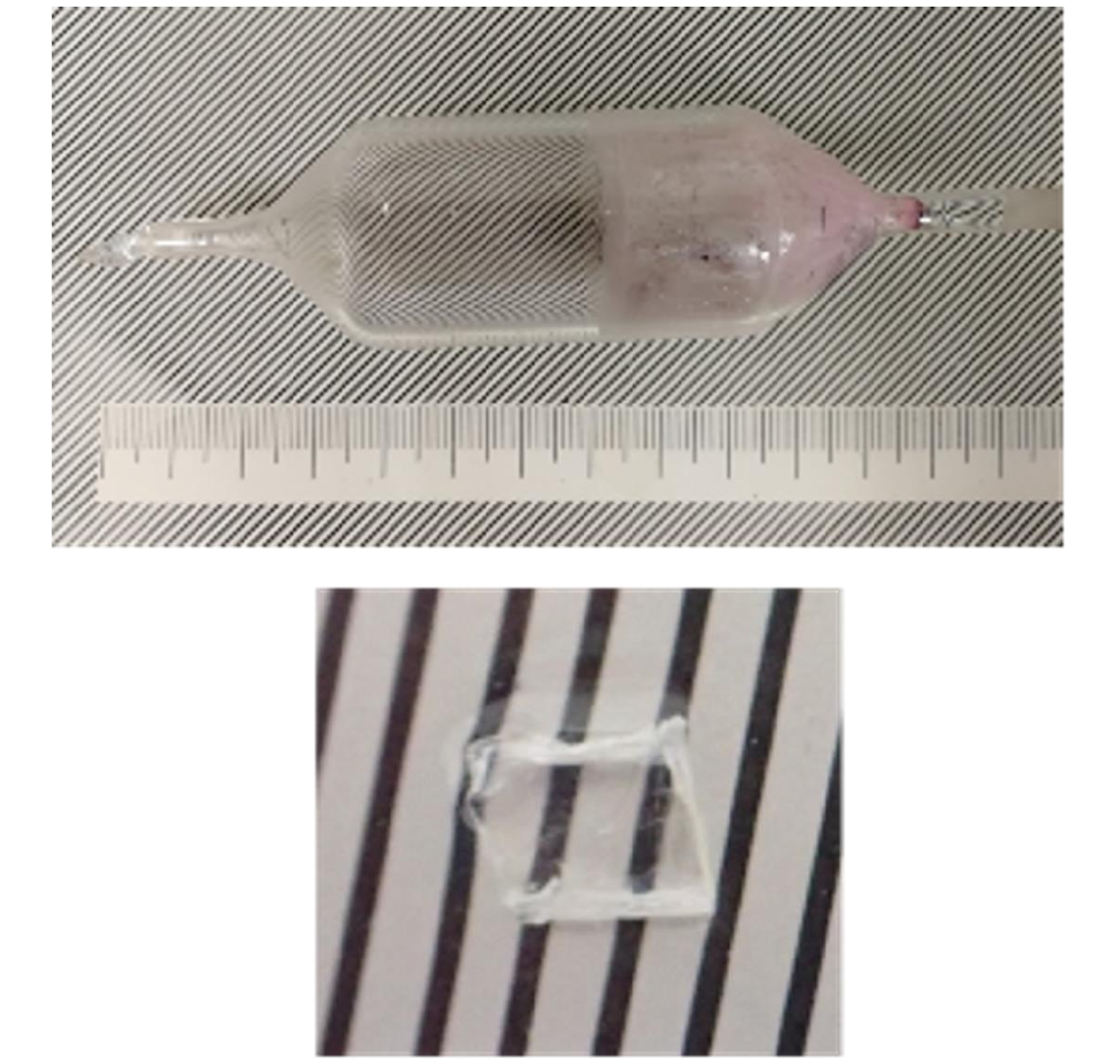}
        \subcaption{CaI$_2$}
        \label{subfig:pic_cai2}
      \end{minipage} &
      \begin{minipage}[t]{0.3\linewidth}
        \centering
        \includegraphics[keepaspectratio, scale=0.3]{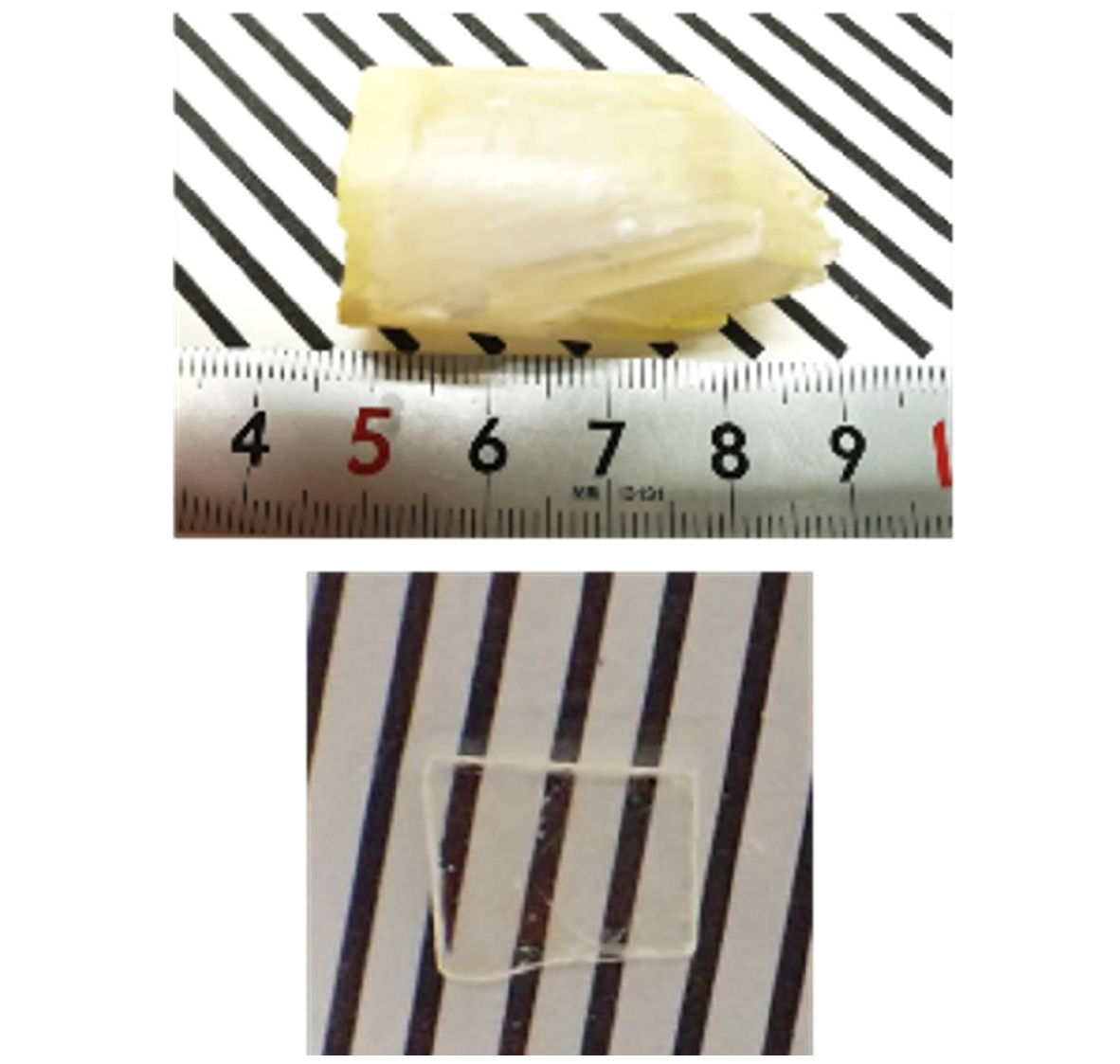}
        \subcaption{CaBrI}
        \label{subfig:pic_cabri}
      \end{minipage} &
      \begin{minipage}[t]{0.3\linewidth}
        \centering
        \includegraphics[keepaspectratio, scale=0.3]{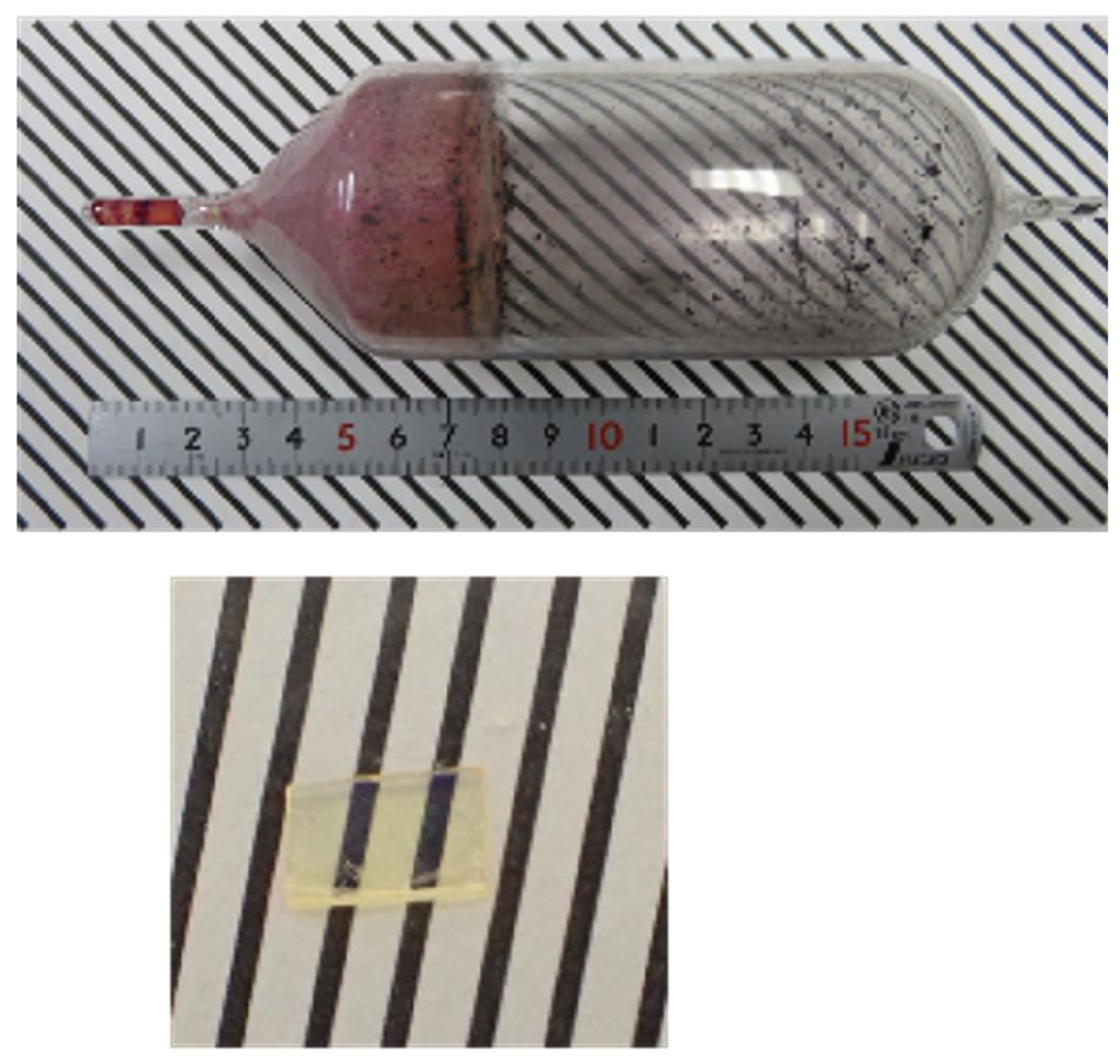}
        \subcaption{CaBrI:Eu}
        \label{subfig:pic_cabrieu}
      \end{minipage}
    \end{tabular}
     \caption{Photographs of grown crystals used in this study. (top) as-grown crystal, (bottom) cut and polished sample.}
     \label{fig:crystal}
  \end{figure*}

\subsection{Crystal growth} \label{ssec:crystal}
CaI$_2$, CaBr$_{0.7}$I$_{1.3}$~(referred to as CaBrI in this paper), and Eu 2 \% doped CaBr$_{0.7}$I$_{1.3}$~(referred to as CaBrI:Eu in this paper) were grown using the vertical Bridgman--Stockbarger method. 
The diameters of the grown crystals were CaI$_2$; 2~inch, CaBrI; 1~inch, and CaBrI:Eu; 2~inch. The starting materials were CaI$_2$, CaBr$_2$, and EuBr$_3$ beads with 4N purity. After these materials were baked at 200°C, they were sealed in quartz ampoules. 
To remove moisture and oxygen from the quartz ampoule, a vacuum of 10$^{-4}$~Pa was drawn, and then the atmosphere was replaced with a mixture of Ar of 9N purity gas and ~1\% SiCl$_4$ gas. 
After the quartz ampoule was sealed, the ampoule was placed in a Pt heater installed in a high-frequency induction coil. The quartz ampoules and Pt heaters were surrounded by alumina insulators to control the temperature gradient along the growth direction. 
The details of the crystal growth setup are described in \cite{CaClBrI}. The quartz ampoule was heated to the melting point of CaI$_2$, CaBrI, and CaBrI:Eu, and then the ampoule was pulled down at a rate of 0.003--0.009 mm/min. 
All the crystals were cut, polished, and measured in a dry room (maintained at below 3\% humidity). 0.5 mm thick transparent samples were obtained. Photographs of the grown crystals and the polished samples are shown in Fig. \ref{fig:crystal}.

% \begin{figure}[ht]
%  \centering
%  \includegraphics[keepaspectratio, width=\linewidth]{figures/crystal_picture.png}
%  \caption[grown crystal]{Photographs of grown crystals used in this study. (top) as-grown crystal, (bottom) cut and polished sample.}
%  \label{fig:crystal}
% \end{figure}

\subsection{Measurement setup and data preparation} \label{ssec:setup}

Photoluminescence (PL) emission and excitation spectra were measured using an Edinburgh Instruments Xe-900 fluorescence spectrometer and Xe lamp excitation. Both the excitation and emission monochromator were set with a 1 nm bandpass slit.

PSD performance was evaluated using radioisotope (RI) sources of $^{241}$Am as the $\alpha$-ray source and $^{137}$Cs as the $\gamma$-ray source. 
The radioactivity levels of $^{241}$Am and $^{137}$Cs were 4~MBq and 10~MBq, respectively. The distance between the RI sources and the samples was 4~mm. 
\textcolor{black}{To suppress the waveform pileup, a tungsten collimator with a thickness of 2~mm and a hole diameter of 2mm$\phi$ was placed between samples and RI sources.}
\textcolor{black}{Scintillation lights were collected using a photomultiplier tube (PMT, R7600-200, Hamamatsu Photonics K.K., Japan). The scintillator samples were connected directly via optical grease (6262A, OHYO KOKEN KOGYO Co., Ltd.) without reflectors, with an operating voltage of 600~V. The signal from the PMT was amplified using an eight-channel PM AMP (RPN-092, HAYASHI-REPIC CO., LTD., Tokyo, Japan).}
The amplified signal was digitized and acquired by the two-channel USB WaveCatcher module \cite{WaveCatcher}. The module had a dynamic range of $\pm$1.25~V coded over 12~bits, bandwidth of 500~MHz, and sampling rate of 400~MHz. The sampling depth of each event was 1,024~samples. In the module, the trigger levels for waveform recording were set to (CaI$_2$, $\alpha$): 50~mV, (CaI$_2$, $\gamma$): 30~mV, (CaBrI, $\alpha$):30~mV, (CaBrI, $\gamma$): 15~mV, (CaBrI:Eu, $\alpha$): 40~mV, and (CaBrI:Eu, $\gamma$): 20~mV. The recording started at around 80~sampling points (200~ns) before the trigger.
As data preparation, the baseline was first determined by calculating the average value of the first 50 sampling points (125~ns) of the waveform. 
The baseline was then subtracted from each waveform. 
For conventional methods (DG and DF methods), the ``starting point'' was defined as the point at which the signal first crossed 20~mV.
% The total area of each waveform was calculated by first finding the baseline from the average of the first 50 sampling points (125 ns). 
% The threshold is set very close to this baseline, much smaller than the height of a signal pulse and higher than most electronic noise. 
The pulse area was then calculated as the sum of the areas of each waveform region that contained at least one sample that had exceeded this threshold. 
The energy was calibrated using the 0.662 MeV peak of $^{137}$Cs. 
To calculate the PSD index value, ``normalized waveforms'' were used, where the waveforms were normalized with respect to the area of each waveform.

\begin{figure}[htbp]
    \centering
    \begin{tabular}{c}
      \begin{minipage}[t]{0.5\textwidth}
        % \centering
        \includegraphics[keepaspectratio, scale=0.45]{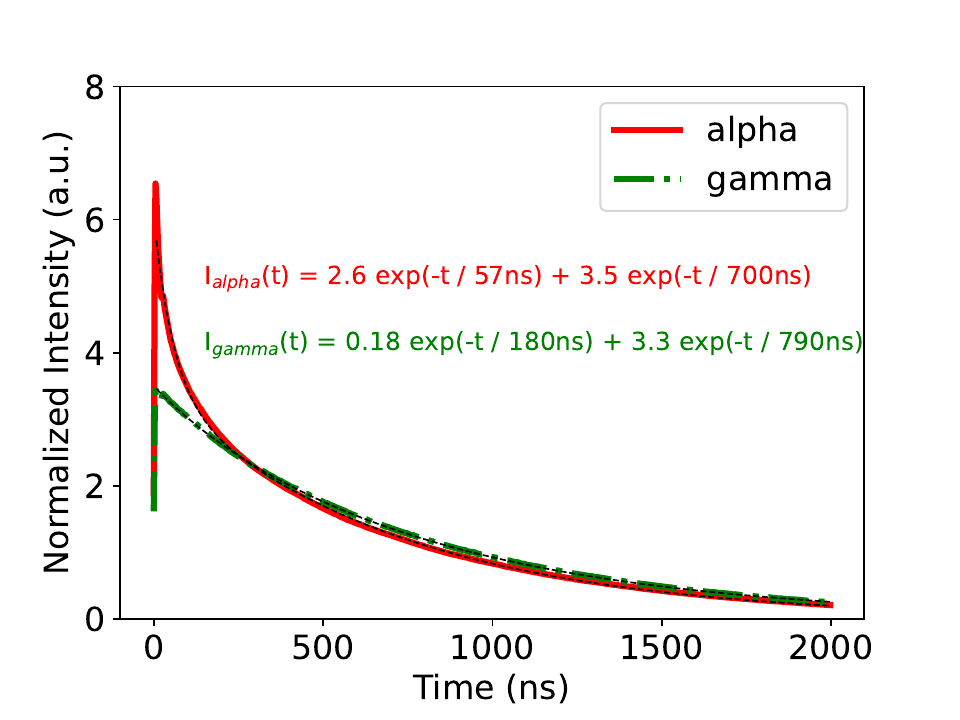}
        \subcaption{CaI$_2$}
        \label{subfig:wf_cai2}
      \end{minipage} \\
      \begin{minipage}[t]{0.5\textwidth}
        % \centering
        \includegraphics[keepaspectratio, scale=0.45]{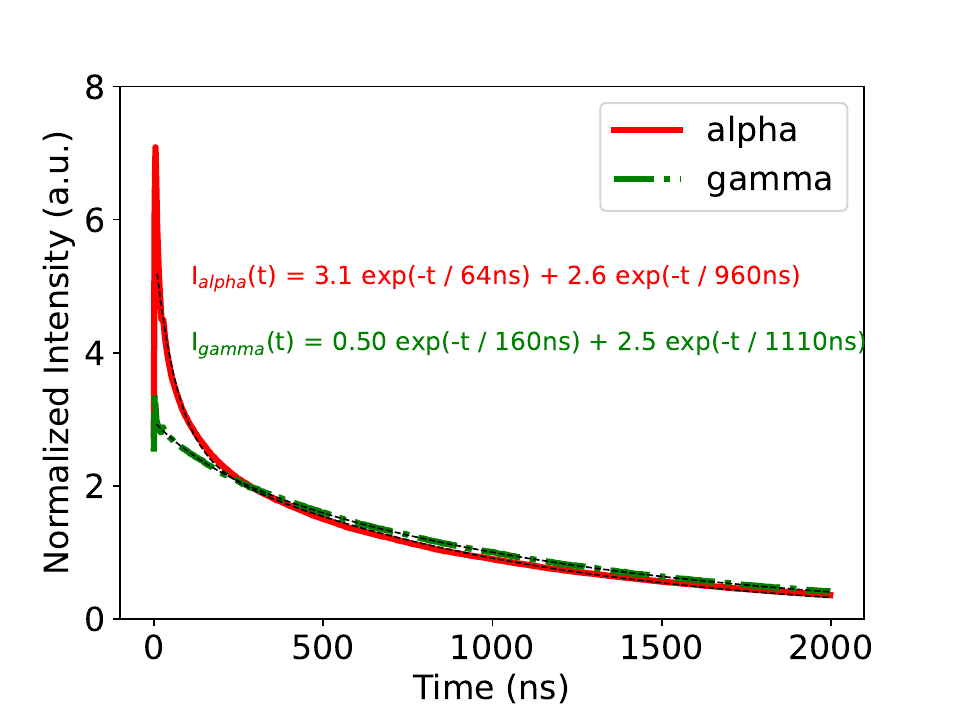}
        \subcaption{CaBrI}
        \label{subfig:wf_cabri}
      \end{minipage} \\
      \begin{minipage}[t]{0.5\textwidth}
        % \centering
        \includegraphics[keepaspectratio, scale=0.45]{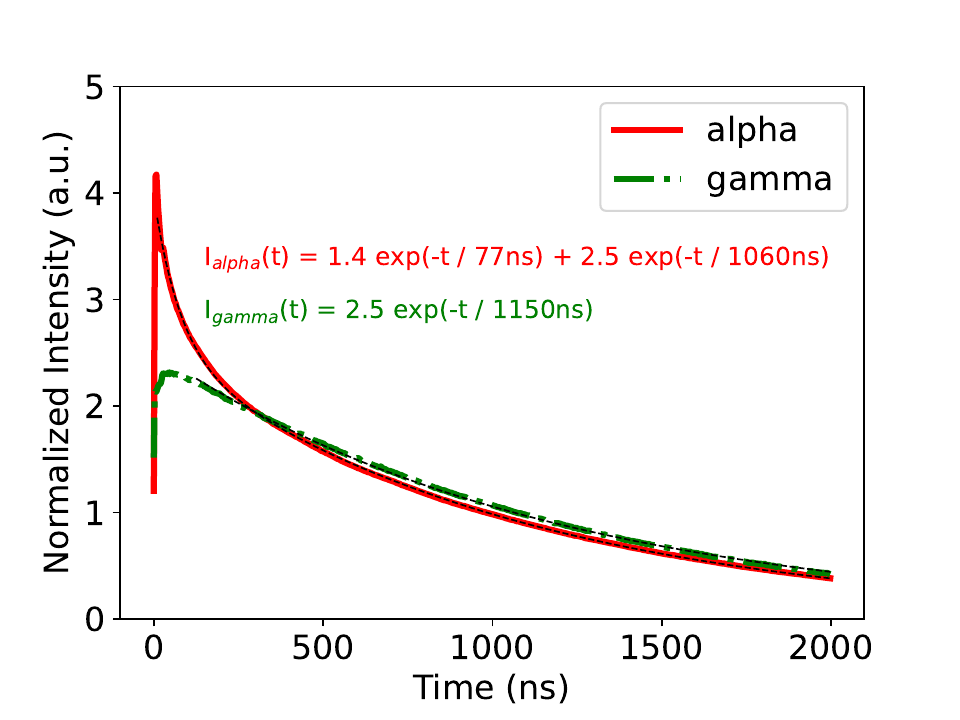}
        \subcaption{CaBrI:Eu}
        \label{subfig:wf_cabrieu}
      \end{minipage}
    \end{tabular}
     \caption{Average waveforms using the three scintillators for $\alpha$ and $\gamma$ sources. The dashed black line represents the fitting functions.}
     \label{fig:waveform}
  \end{figure}

\begin{table}[htbp]
  \centering
  \caption{Fitting results of the decays for CaI$_2$, CaBrI, and CaBrI:Eu.}
  \begin{tabular}{lcrr}
    Material    &   Source       &   Primary decay   &   Secondary decay \\ \hline
    CaI$_2$     &   $^{241}$Am   &   57 ns (6\%)     &   700 ns (94\%)  \\
                &   $^{137}$Cs   &   180 ns (1\%)    &   790 ns (99\%)  \\ 
    CaBrI       &   $^{241}$Am   &   64 ns (7\%)     &   960 ns (93\%)  \\
                &   $^{137}$Cs   &   160 ns (3\%)    &   1110 ns (97\%)  \\ 
    CaBrI:Eu    &   $^{241}$Am   &   77 ns (4\%)     &   1060 ns (94\%) \\
                &   $^{137}$Cs   &   - (0\%)         &   1150 ns (100\%)  \\ \hline
  \end{tabular}
  \label{tb:wf-fit}
\end{table}

The average pulse shapes for the  $\alpha$- and $\gamma$-rays measured using CaI$_2$, CaBrI, and CaBrI:Eu are shown in Fig. \ref{fig:waveform}. All waveforms were normalized by their area. 
The decay time of the average waveform was calculated by fitting one or the sum of two exponentials:
\begin{displaymath}
 A(t) = A_1 \exp(-t/\tau_1) + A_2 \exp(-t/\tau_2),
\end{displaymath}
where $A_1$ and $A_2$, and $\tau_1$ and $\tau_2$ are the amplitudes and decay times of the waveforms, respectively. The fitting ranges started from (CaI$_2$, $\alpha$): 7.5~ns, (CaI$_2$, $\gamma$): 5~ns, (CaBrI, $\alpha$): 10~ns, (CaBrI, $\gamma$): 7.5~ns, (CaBrI:Eu, $\alpha$): 10~ns, and (CaBrI:Eu, $\gamma$): 125~ns after the ''starting point.''
All pulse shapes were successfully fitted by one or the sum of two exponential functions.
The fitting values are summarized in Table. \ref{tb:wf-fit}. For all samples, the pulse shapes and decays of the $\alpha$- and $\gamma$-rays were completely different. The primary decay ratios obtained from the fitting were very low for all samples.
The $\gamma$-ray response of CaBrI:Eu was different from that of the other undoped samples. This may be because of the change in the scintillation process of the scintillator when Eu doping was used.

Eighty thousand waveforms were recorded as gamma ray events in the CaBrI sample. As other events, 50,000 waveforms were recorded. 
\textcolor{black}{
$^{241}$Am decays mainly via alpha decay, with a weak $\gamma$-ray byproduct. The alpha decay is shown as follows:
\begin{displaymath}
 ^{241}_{95}\mathrm{Am} \xrightarrow{432.2y}~^{237}_{93}\mathrm{Np} + ^{4}_{2}\alpha + \gamma ~59.5409~keV,
\end{displaymath}
The $\gamma$-ray energy is 59.5409~keV for the most part (35.9\%), with little amounts of other energies such as 26.3~keV (2.4\%), and 33.2~keV (0.13\%). The $\gamma$-ray energy over 60~keV is less than 0.06\% of the time.
To exclude the events of the 59.5 keV $\gamma$-rays emitted from $^{241}$Am from the analysis, only events with energies above 100 \textcolor{black}{keVee} were selected for PSD analysis.
}
 In this study, the PSD performance in the low-energy region was also evaluated. For that analysis, events with energies between 100 and 200 \textcolor{black}{keVee} were selected. The number of events in these datasets is summarized in Table \ref{tb:data}.

\begin{table}[htbp]
  \centering
  \caption{Number of events in the dataset prepared for analysis}
  \begin{tabular}{lcrrr}
    Materials & Source & Total & $>$100 \textcolor{black}{keVee} & 100--200 \textcolor{black}{keVee}\\ \hline \hline
    CaI$_2$ & $^{241}$Am & 50000 & 38349 & 6382 \\
     & $^{137}$Cs & 50000 & 26748 & 12895 \\
    CaBrI & $^{241}$Am & 50000 & 46686 & 4804 \\
     & $^{137}$Cs & 80000 & 28306 & 13292 \\
    CaBrI:Eu & $^{241}$Am & 50000 & 48251 & 1709 \\
     & $^{137}$Cs & 50000 & 26686 & 12304 \\ \hline
  \end{tabular}
  \label{tb:data}
\end{table}

%% file: 2_2_method.tex
\subsection{Analysis methods} \label{ssec:analysis}

The PSD performances for these three crystals were evaluated using several methods. 
The first was the DG method and the second was a DF method. 
These two methods have been widely used for the PSD of scintillators \cite{TAMAGAWA2015192, Mizukoshi_2019}. 
The alternative methods used were based on machine learning. The MLP %network
and CNN were used as a deep network and compared with the conventional methods quantitatively.
Each dataset was randomly divided into three subsets containing 70\%, 20\%, and 10\% of the data as training, validation, and test datasets, respectively.

The F1 score, which is defined as the harmonic mean of Precision and Recall, was introduced as an evaluation function for PSD capability. All events were defined as follows:

\begin{table}[hbtp]
%   \caption{}
%   \label{table:data_type}
  \centering
  \begin{tabular}{ccc}
       & \multicolumn{2}{c}{\textcolor{black}{Prediction}} \\
    \textcolor{black}{Observation} & $\alpha$ & $\gamma$ \\
    \hline
     $\alpha$  & True positive (TP) & False negative (FN) \\
     $\gamma$  & False positive (FP) & True negative (TN) \\
    \hline
  \end{tabular}
\end{table}

Then, the Precision and Recall were defined as

\begin{displaymath}
 \mbox{Precision} = \frac{\mbox{TP}}{\mbox{TP + FP}},
\end{displaymath}
\begin{displaymath}
 \mbox{Recall} = \frac{\mbox{TP}}{\mbox{TP + FN}},
\end{displaymath}

and the F1 score was obtained as

%\begin{displaymath}
% \mbox{F-measure} = \frac{2 %\times \mbox{Recall} \times %\mbox{Precision}}{\mbox{Recal%l} + \mbox{Precision}}
%\end{displaymath}

\begin{eqnarray}
\label{eq:f1}
 \mbox{F1} &=& \frac{2 \times \mbox{Recall} \times \mbox{Precision}}{\mbox{Recall} + \mbox{Precision}} \nonumber \\
&=& \frac{2 \mathrm{TP}}{\mathrm{TP}+\mathrm{FP}+N_1},
\end{eqnarray}

where $N_1=\mathrm{TP}+\mathrm{FN}$, which is a constant because it is the total number of events in the $\alpha$-source data.
The PSD separation criteria for each method were set to achieve the highest F1 score using the validation data for each crystal.
The closer the F1 score is to 1, the better the separation of $\alpha$ and $\gamma$.

Only statistical errors were taken into account for the error of the F1 score.
As shown in Eq.~(1), the F1 score is expressed in terms of two variables, True positive~(TP) and False positive~FP.
TP and FP are independent variables because they are provided by different datasets;
that is, the error of the F1 score $\Delta \mathrm{F1}$ can be calculated from $\Delta \mathrm{TP}$ and $\Delta \mathrm{FP}$ using an error propagation expressed as follows:

\begin{equation}
 (\Delta \mathrm{F1})^2 =  \left(\frac{\partial \mathrm{F1}}{\partial \mathrm{TP}}\right)^2\Delta \mathrm{TP}^2 + \left(\frac{\partial \mathrm{F1}}{\partial \mathrm{FP}}\right)^2 \Delta \mathrm{FP}^2. 
\end{equation}

The derivatives of the F1 score are 
\begin{equation}
\begin{split}
\frac{\partial \mathrm{F1}}{\partial \mathrm{FP}} = \frac{-2\mathrm{TP}}{(\mathrm{FP}+\mathrm{TP}+N_1)^2}, \\
\frac{\partial \mathrm{F1}}{\partial \mathrm{TP}} = \frac{2(\mathrm{FP}+N_1)}{(\mathrm{FP}+\mathrm{TP}+N_1)^2} 
\end{split}
\end{equation}
. 
The statistical errors of TP and FP can be obtained from the binomial distribution as follows:
\begin{equation}
\begin{split}
\Delta \mathrm{TP} &= \sqrt{\mathrm{TP}\left(1-\frac{\mathrm{TP}}{N_1}\right)}, \\
\Delta \mathrm{FP} &= \sqrt{\mathrm{FP}\left(1-\frac{\mathrm{FP}}{N_2}\right)}, 
\end{split}
\end{equation}

where $N_2 = \mathrm{FP}+\mathrm{TN}$, which is the number of total events in the $\gamma$-source data.

%The statistical error for the F1 score is expressed by the following equation.
%\begin{eqnarray}
% \Delta F = \sqrt{\left(\frac{2(\mathrm{TP}+FP+FN)}{2\mathrm{TP}+FP+FN}\right)^2\left(\mathrm{TP}\left(1-\frac{\mathrm{TP}}{\mathrm{TP}+FN}\right)\right)  \nonumber \\ 
% \overline{+ \left(\frac{2\mathrm{TP}}{(2\mathrm{TP}+FP+FN)^2}\right)^2\left(FP\left(1-\frac{FP}{FP+TN}\right)\right)}}
%\end{eqnarray}

\begin{figure*}[h]
    \begin{tabular}{cc}
      \begin{minipage}[t]{0.45\linewidth}
        \centering
        \includegraphics[keepaspectratio, scale=0.55]{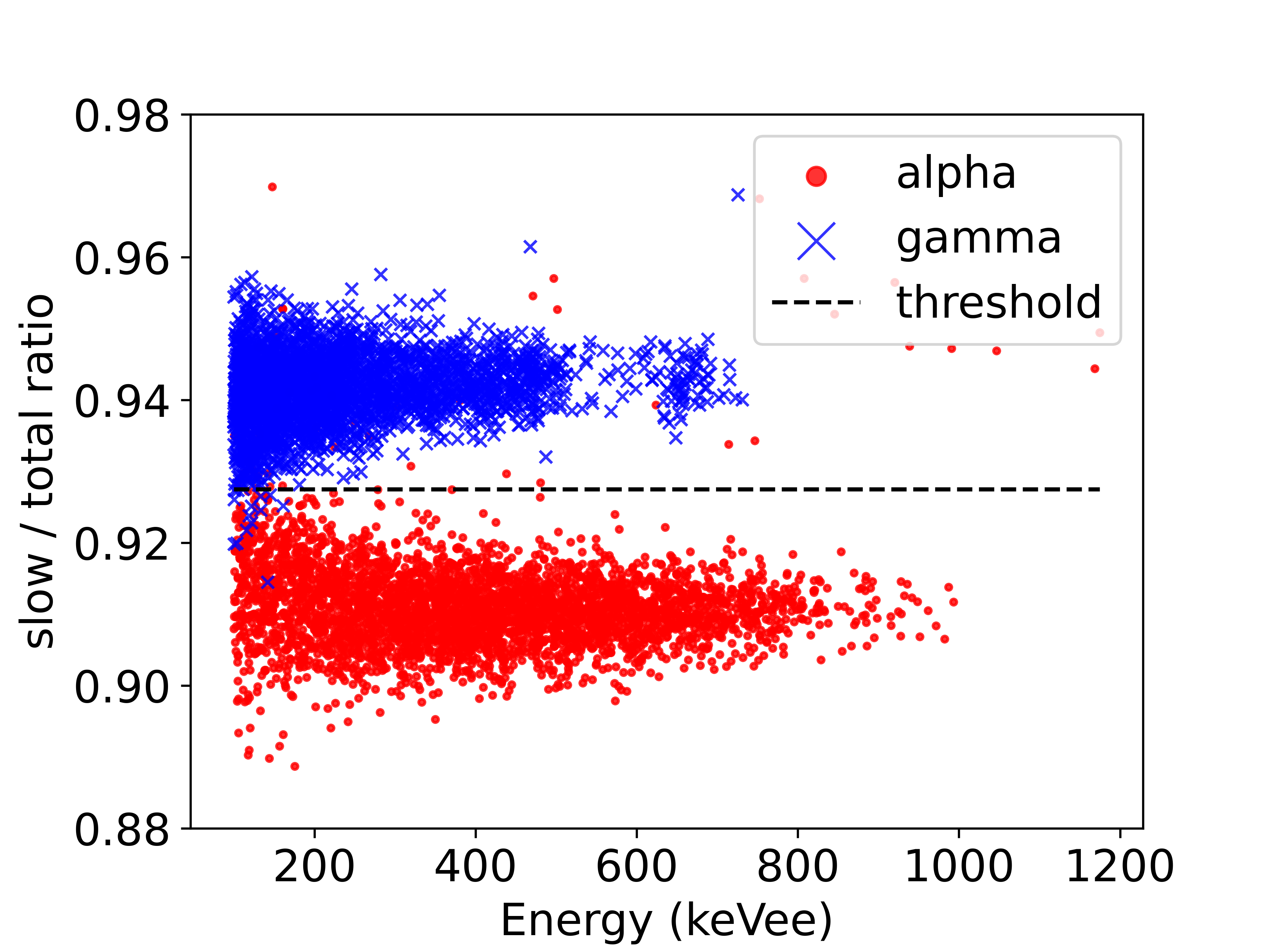}
        \subcaption{Double gate method}
        \label{subfig:dg}
      \end{minipage} &
      \begin{minipage}[t]{0.45\linewidth}
        \centering
        \includegraphics[keepaspectratio, scale=0.55]{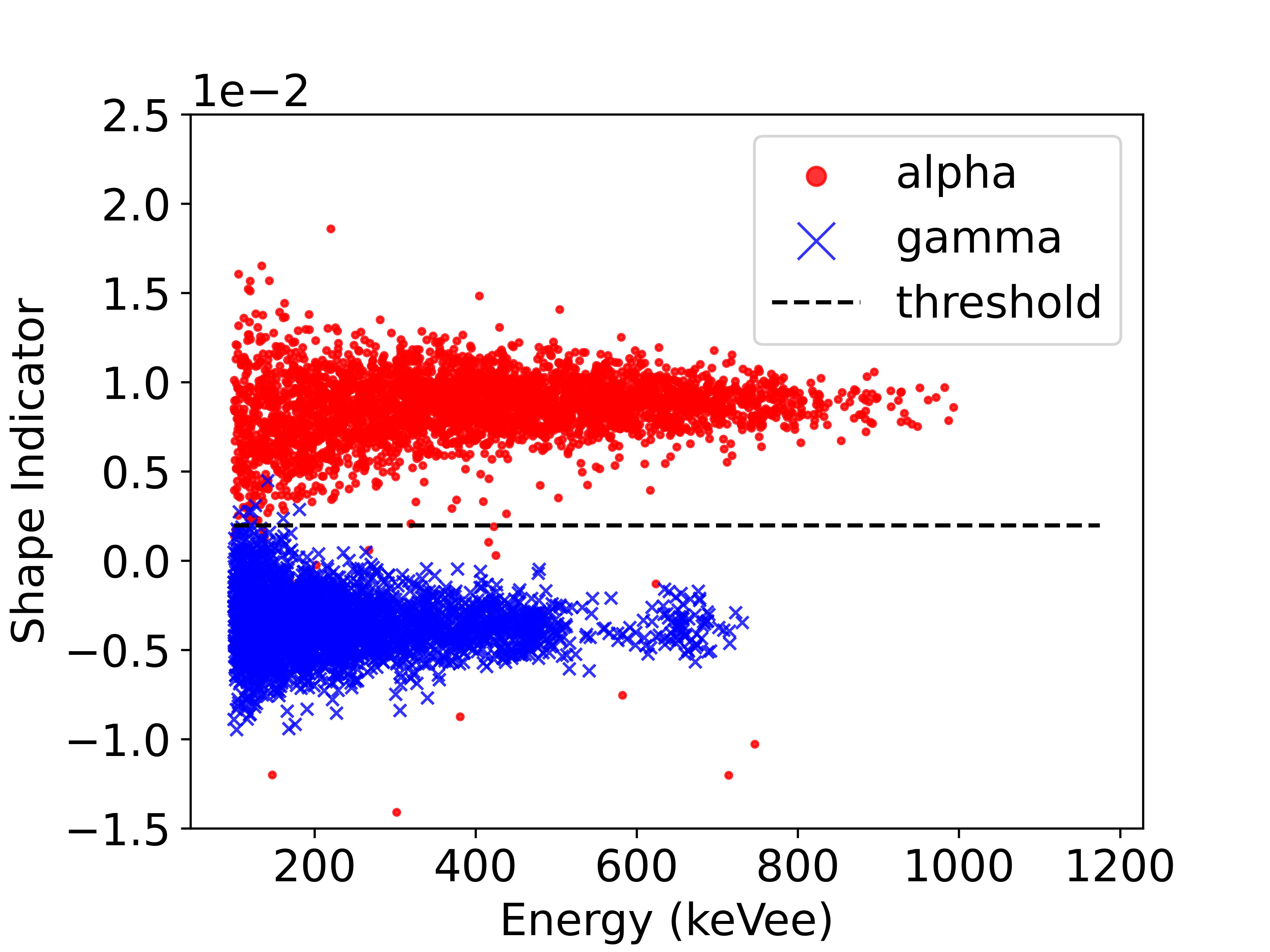}
        \subcaption{Digital filter method}
        \label{subfig:si}
      \end{minipage} \\
   
      \begin{minipage}[t]{0.45\linewidth}
        \centering
        \includegraphics[keepaspectratio, scale=0.55]{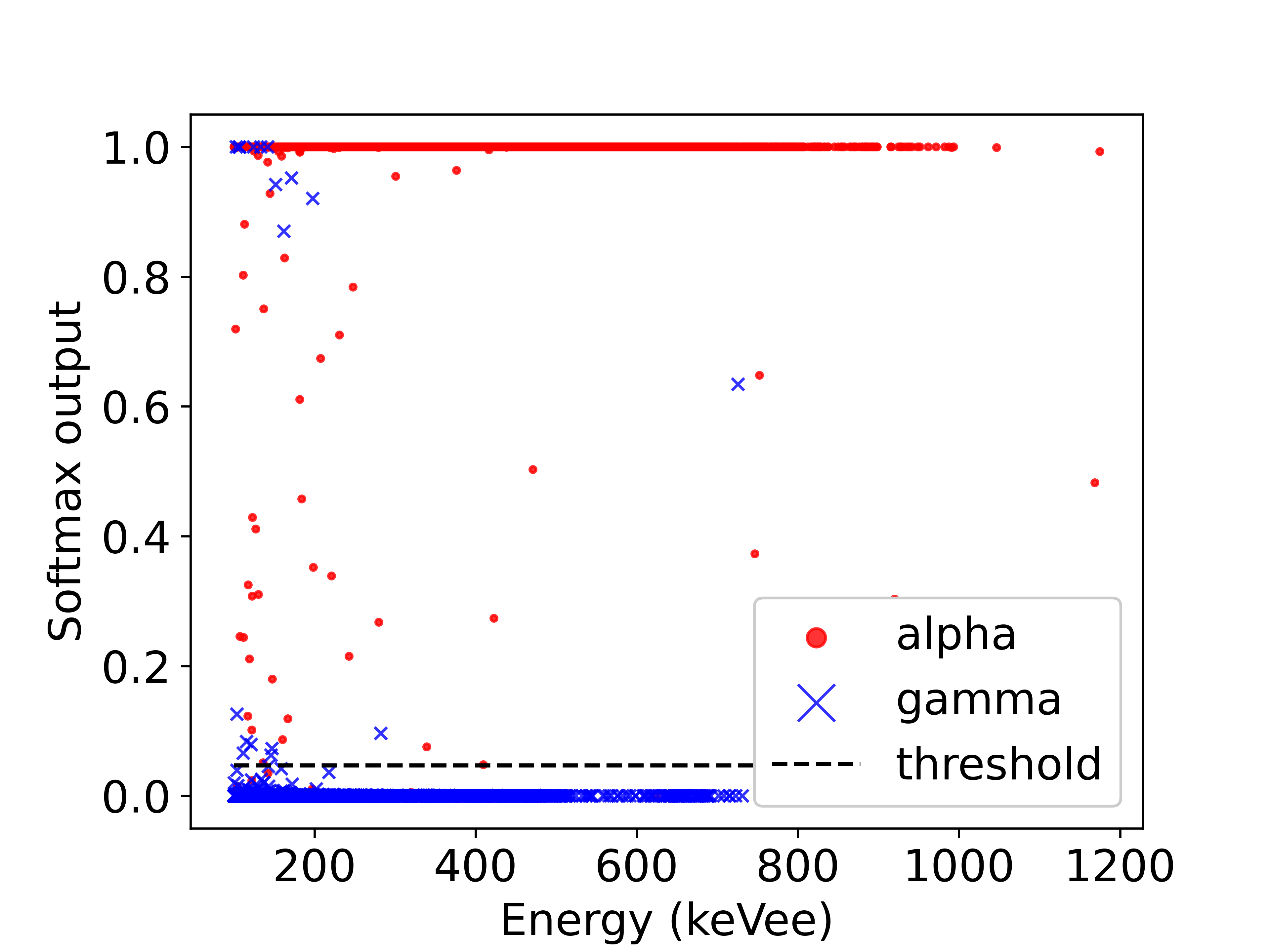}
        \subcaption{Multilayer perceptron %network
        }
        \label{subfig:mlp}
      \end{minipage} &
      \begin{minipage}[t]{0.45\linewidth}
        \centering
        \includegraphics[keepaspectratio, scale=0.55]{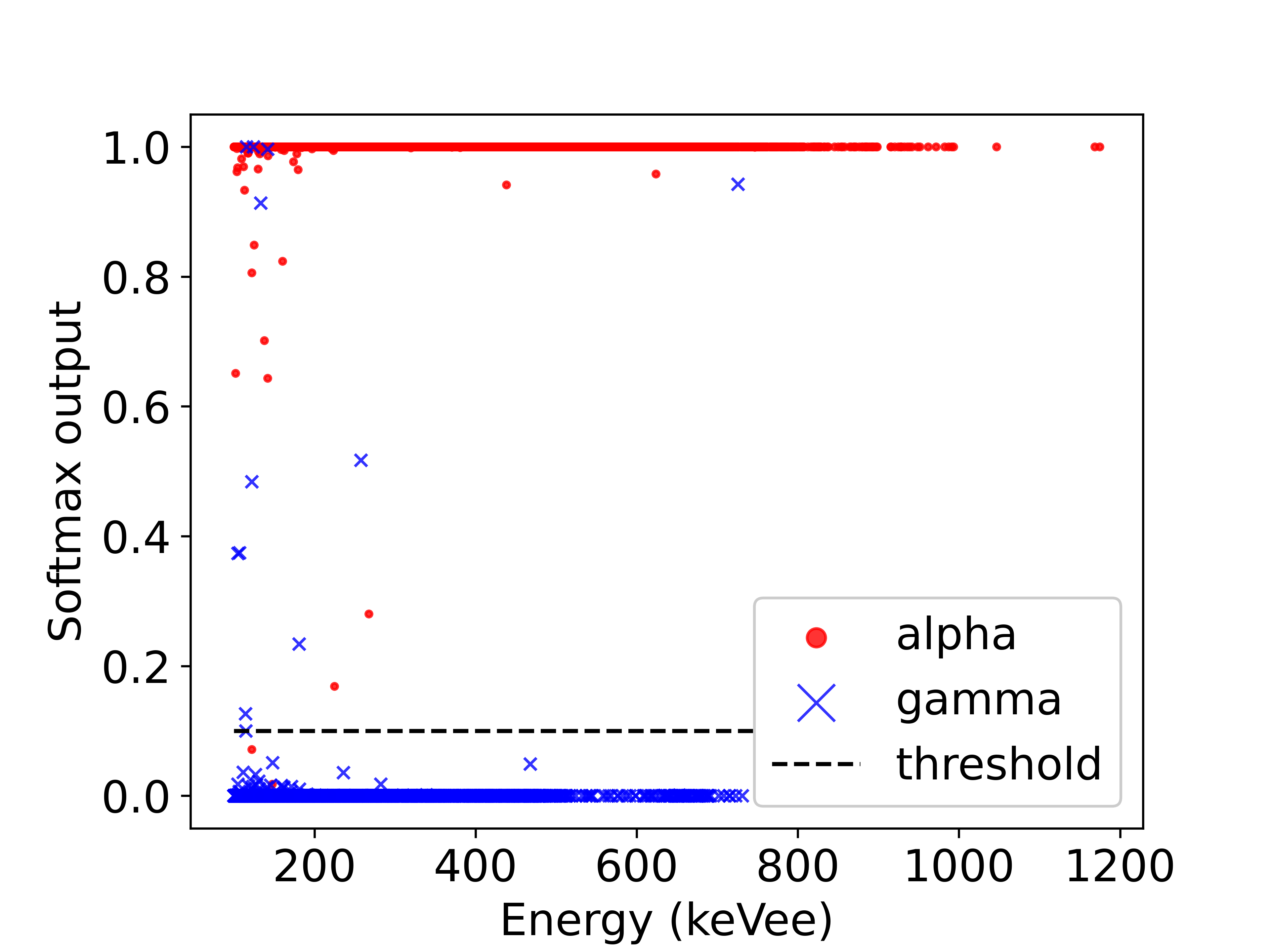}
        \subcaption{Convolutional neural network}
        \label{subfig:cnn}
      \end{minipage} 
    \end{tabular}
     \caption{PSD scatter plot vs electron equivalent energy for CaI$_2$ test data.}
     \label{fig:psd}
  \end{figure*}

\subsubsection{Double gate method: DG} \label{sssec:dual}
The DG method is widely used for PSD.
A DG method defines two gates that integrate pulses: one gate that includes the entire pulse (called `Total') and another gate that excludes the initial part of the pulse (called `Slow'). 
The pulse is generally composed of several components that have different decay constants. 
The ratio of the components depends on the incoming particles.
Thus, the ratio, `Slow'/`Total' can distinguish the incoming particle.
This method is easy to implement, not only for offline analysis but also online or hardware analysis~\cite{dpp_psd}.

In the DG method used in this study, the training data were used to optimize the starting point of the slow integration. The best PSD separation was observed in the slow integration region of 45-2000~ns, 50-2000~ns, and 70-2000~ns for CaI$_2$, CaBrI, and CaBrI:Eu, respectively. Then, the PSD separation criterion was set at a Slow/Total value of 0.9275, 0.9233, and 0.9206 for CaI$_2$, CaBrI, and CaBrI:Eu, respectively, using validation data.
The best Slow/Total scatter plot and separation criterion for CaI$_2$ is shown in Fig. \ref{subfig:dg} as a function of the electron equivalent energy.

\subsubsection{Digital filter method: DF} \label{sssec:si}
In DF methods, the waveform of each event is compared with the waveform of the reference $\alpha$ and $\gamma$ radiation. In this study, a shape indicator (SI) \cite{TAMAGAWA2015192,Mizukoshi_2019} was used as a DF method. The formula for the SI applied to each waveform is
\begin{displaymath}
 \mathrm{SI} = \frac{\Sigma f(t_k)P(t_k)}{\Sigma f(t_k)},
\end{displaymath}
where $f(t_k)$ is the digitized amplitude of each waveform (at time $t_k$). The weight function $P(t)$ is defined as
\begin{displaymath}
 P(t_k) = \frac{f_\alpha(t_k) - f_\gamma(t_k)}{f_\alpha(t_k) + f_\gamma(t_k)},
\end{displaymath}
where $f_\alpha(t_k)$ and $f_\gamma(t_k)$ are the reference pulse shapes of the $\alpha$-ray and $\gamma$-ray, respectively.
In this study, the average pulse waveforms of the training data for $\alpha$ and $\gamma$ were applied to the reference $f_\alpha(t_k)$ and $f_\gamma(t_k)$.

For the SI used in this study, the training data were used to calculate the weight function $P(t)$. Then, the PSD separation criteria were determined to be SI values of 0.0020, 0.0044, and 0.0035 for CaI$_2$, CaBrI, and CaBrI:Eu, respectively, using the validation data for each crystal.
 The best SI scatter plot and separation criterion for CaI$_2$ are shown in Fig. \ref{subfig:si} as a function of the electron equivalent energy.

%% file: 2_3_ml.tex
\subsection{Machine learning} \label{ssec:ml}

\subsubsection{Multilayer perceptron%network
: MLP}

\begin{figure}[h]
 \centering
 \includegraphics[keepaspectratio,width=\linewidth]{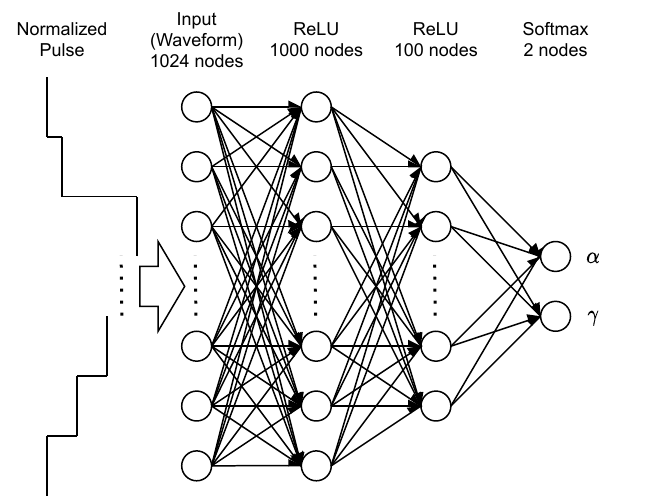}
 \caption{Design of the MLP.}
 \label{fig:ann_design}
\end{figure}

\begin{figure}[t] \centering
\begin{tabular}{c}
\includegraphics[width=2.8in]{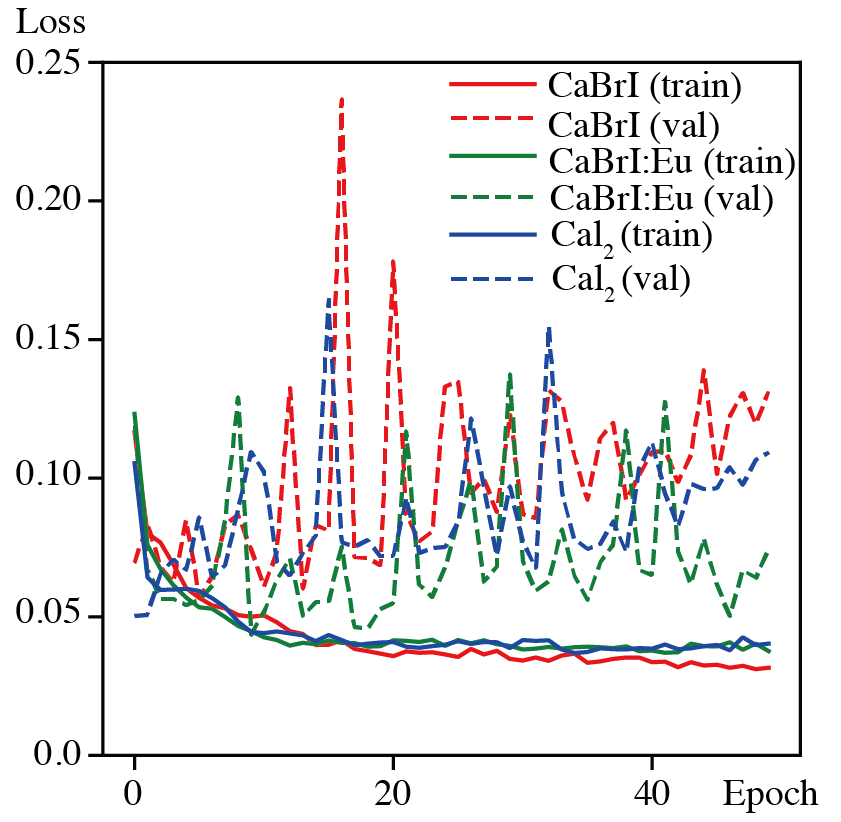}\\
(a)~MLP \\
\includegraphics[width=2.8in]{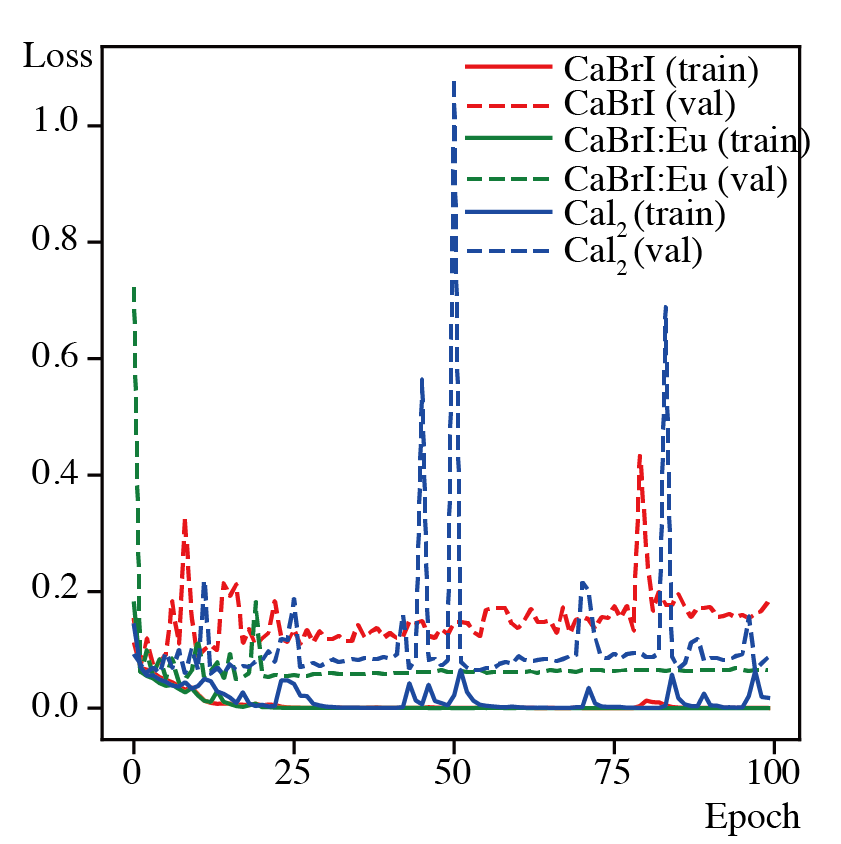}\\
(b)~CNN
\end{tabular}
\caption{Loss values \textcolor{black}{on train and validation data}} \label{fig:loss_mlp_cnn}
\end{figure}

An MLP %network
is one of the simplest artificial neural network.
An MLP has three or more layers of nodes, an input layer, hidden layer(s), and an output layer.
In this study, an MLP that had two hidden layers was implemented with TensorFlow version 2.2.0~(Fig. \ref{fig:ann_design}).
The optimizer was RMSprop with TensorFlow default parameters and the loss function was categorical cross-entropy.
\textcolor{black}{Fig.~\ref{fig:loss_mlp_cnn}~(a) shows the loss values in training.}
The normalised waveform is given to the node in each bin of the input layer.
The activation function of the hidden layer was a rectified linear unit (ReLU). 
The first and second layers had 1,000 and 100 nodes, respectively.
The softmax function was used in the output nodes.
Each output value is in the interval [0,1], so that it can be interpreted as the probability of $\alpha$ and $\gamma$ particle prediction.
In this study, the value provided by a node for the $\alpha$ particle was used as a PSD parameter.
The MLPs were trained using the training data defined in Subsection~\ref{ssec:analysis}.
The validation data were used to ensure that the training process did not over-fit the training data, and determine the best division thresholds of the F1 score.
The separation plot for the test data is shown in Fig.~\ref{subfig:mlp}.
All processes were performed without using test data.

\subsubsection{Convolutional neural network: CNN} 
A CNN model was also used for the PSD. Specifically, the ResNet-18 model~\cite{bib:Miyazaki:ResNet} was adopted, which is widely used in the field of machine learning. Table~\ref{tab:miyazaki:resnet} shows the configuration of ResNet-18 and the output size of the components. Note that batch normalization~\cite{bib:Miyazaki:batch_norm} and the ReLU followed the convolution layers. The residual block retained the input signal of the block and then added that signal to the output of the block. This procedure enabled stable training, even when the number of layers was large. \textcolor{black}{We uses the binary cross entropy to train the mode. The loss values are shown in Fig.~\ref{fig:loss_mlp_cnn}~(b).} \textcolor{black}{The validation losses were oscillated comparing to the training. Generally, these oscillations are caused by overfitting to the training data. Thus, we can keep the best model over the validation losses to avoid the overfitting.} The convolution layers were expected to capture the one-dimensional changes of waveforms. The initial size and channel of a waveform were 1024 and 1, respectively. ResNet-18 gradually expanded the channel by reducing the size. Finally, MLP was used to predict the probability of $\alpha$. The separation plot for the test data is shown in Fig.~\ref{subfig:cnn}.

\begin{table}[t] \centering
\caption{Configuration of the CNN model and its output} \label{tab:miyazaki:resnet}
\begin{tabular}{llrr}
Layer name &Layer configuration &Size & Channel \\ \hline
Input &- &1024 &1\\
Conv1 &Kernel 7, stride 2 & 512 & 64\\
Max pool &Kernel 3, stride 2 &256 &64\\
Layer1 &Residual block $\times$ 2 &256 &64\\
Layer2 &Residual block $\times$ 2 &128 &128\\
Layer3 &Residual block $\times$ 2 &64 &256\\
Layer4 &Residual block $\times$ 2 &32 &512\\
Average pool & - &1 &512\\ 
MLP & - &1 &1 \\ \hline
\end{tabular}
\end{table}

%% file: 3_result.tex
\section{Results and discussion}
\label{sec:result}
\subsection{Photoluminescence and scintillation properties}

\begin{figure}[t]
 \centering
 \includegraphics[keepaspectratio, width=\linewidth]{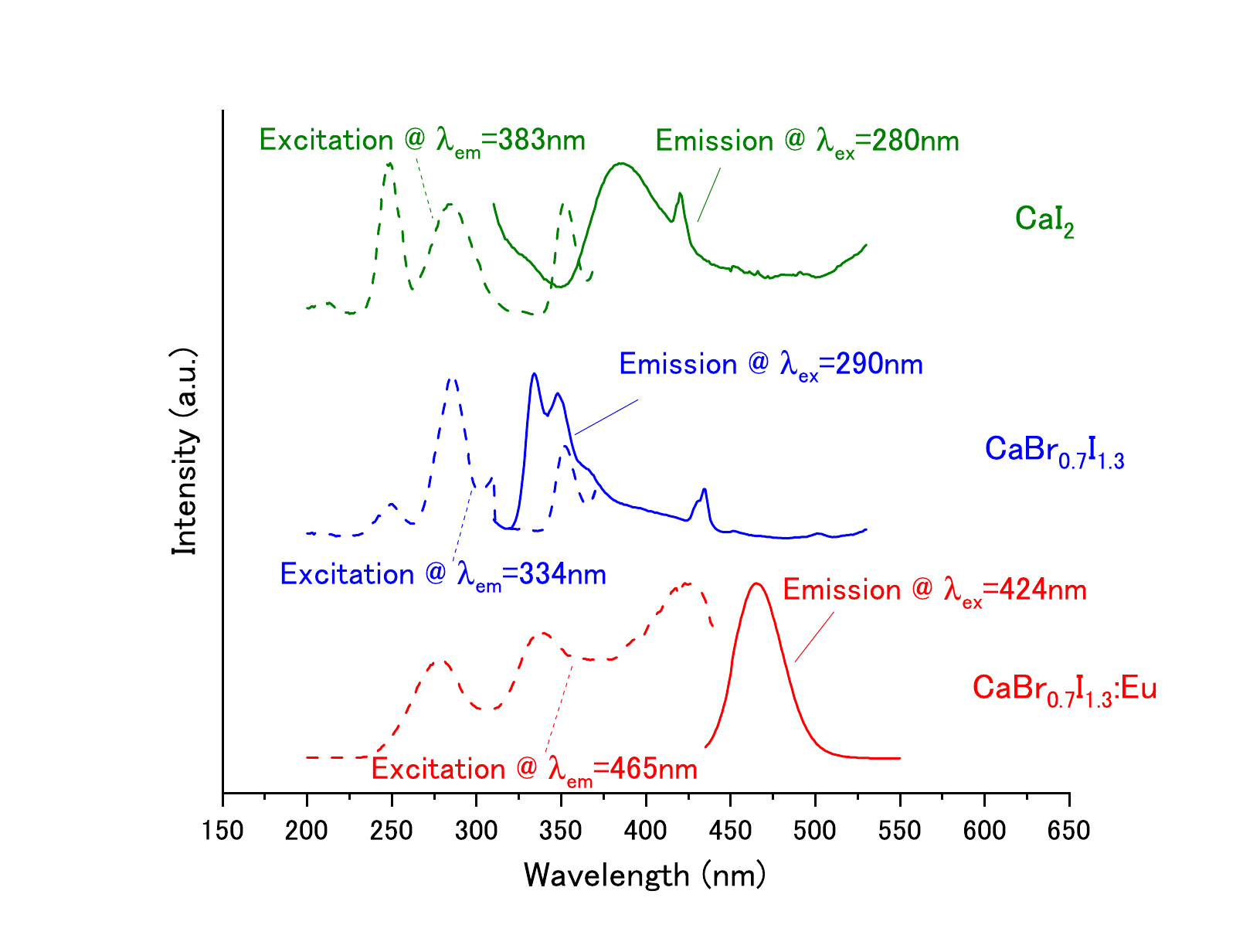}
 \caption[Photoluminescence]{Photoluminescence spectra of the CaI$_2$, CaBrI, and CaBrI:Eu samples shown side by side vertically.}
 \label{fig:pl}
\end{figure}

PL was measured to confirm the presence of Eu dopant. 
In Fig. \ref{fig:pl}, the PL excitation and emission spectra of the CaI$_2$, CaBrI, and CaBrI:Eu crystals are shown side by side vertically.
Only the PL excitation and emission spectra of the CaBrI:Eu sample were shifted to the longer wavelength side.
For the CaBrI:Eu sample, the 4f-5d emission band of Eu$^{2+}$ centered at 465 nm was observed at 365 nm excitation. 
The narrow peaks obserbed in the emission spectra of CaI$_2$ (\textasciitilde 420~nm) and CaBrI (\textasciitilde 430~nm) were most probably caused by the deliquescence of the samples.

\begin{figure}[t]
 \centering
 \includegraphics[keepaspectratio, scale=0.55]{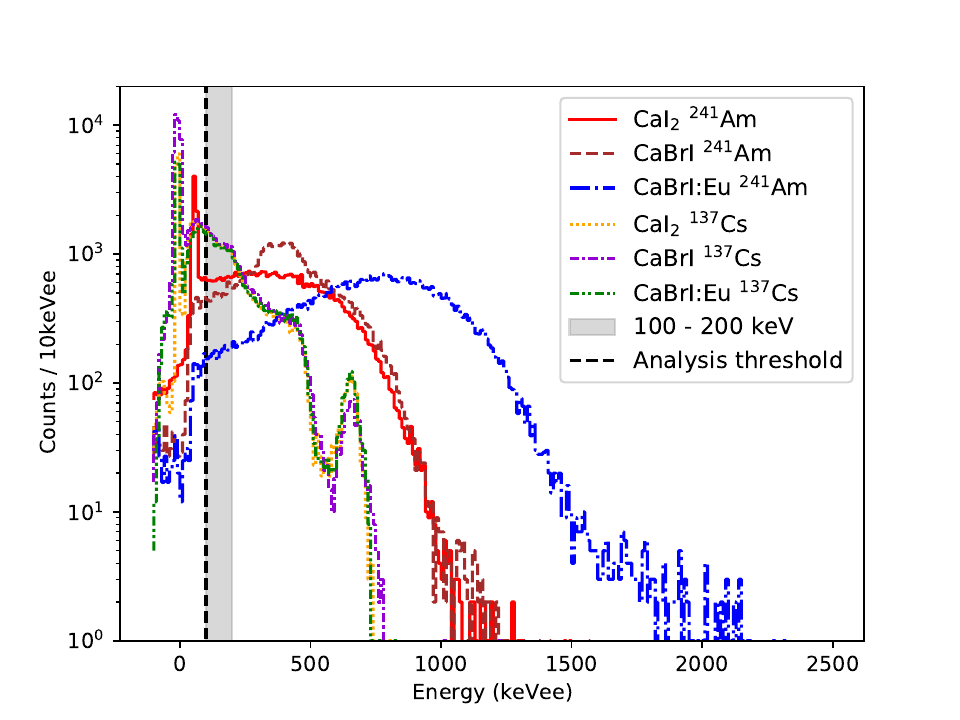}
 \caption[Energy spectra]{Energy spectra using the three scintillators for $\alpha$ and $\gamma$ sources. The black dotted line represents the threshold for analysis. The energy was calibrated using the 0.662 MeV peak of $^{137}$Cs.}
 \label{fig:spectra}
\end{figure}

The calibrated energy spectra are shown in Fig. \ref{fig:spectra}. 
\textcolor{black}{The energy of each event is obtained from the integral of the waveform after subtracting the baseline.}
\textcolor{black}{The measured energy is given in units of electron-equivalent-energy (e.g. keVee or MeVee); 1 MeVee gamma-ray produces 1 MeVee of electron-induced scintillation light, but the heavier particle such as $\alpha$-ray produces less light. The energy was calibrated using the 0.662 MeV peak of $^{137}$Cs.}
The analysis threshold (set to 100 keVee) and the 100--200 keVee region used for low-energy analysis are also shown. 
The light outputs of CaBrI and CaBrI:Eu were calculated assuming a light output of 107,000 ph/MeV for the CaI$_2$ sample.
\textcolor{black}{The positions of the respective photoabsorption peaks were compared, and the light outputs of CaI$_2$, CaBrI, and CaBrI:Eu were estimated to be 107,000, 51,000, and 79,000 ph/MeV, respectively. The $\alpha$ sources have been observed as continuous spectra because the $\alpha$ particles are attenuated by the air or aluminium film between the external source and each scintillator. We used the calibrated energy with the electron equivalent~(keVee).}

\textcolor{black}{ It is well known that, for inorganic and organic scintillators, heavy particles, which lead to a higher ionization density than a $\gamma$-ray, for example, an $\alpha$-ray, have even lower scintillation efficiency and a smaller scintillation intensity than that produced by a $\gamma$-ray of the same energy \cite{birks1951}. The $\alpha$/$\gamma$ ratio, defined as the ratio of the output to the $\alpha$- and $\gamma$-rays of the same energy, is different for each scintillator. Compared with non-doped CaI$_2$ and CaBrI, the CaBrI:Eu sample had a higher $\alpha$/$\gamma$ ratio. }

\subsection{Pulse shape discrimination}
\subsubsection{Comparison of the four methods}
The PSD performance for the three crystals was evaluated using four analysis methods: (1) DG method, (2) DF method, (3) MLP, and (4) CNN.

\begin{figure}[ht]
\centering
\includegraphics[keepaspectratio, width=\linewidth]
{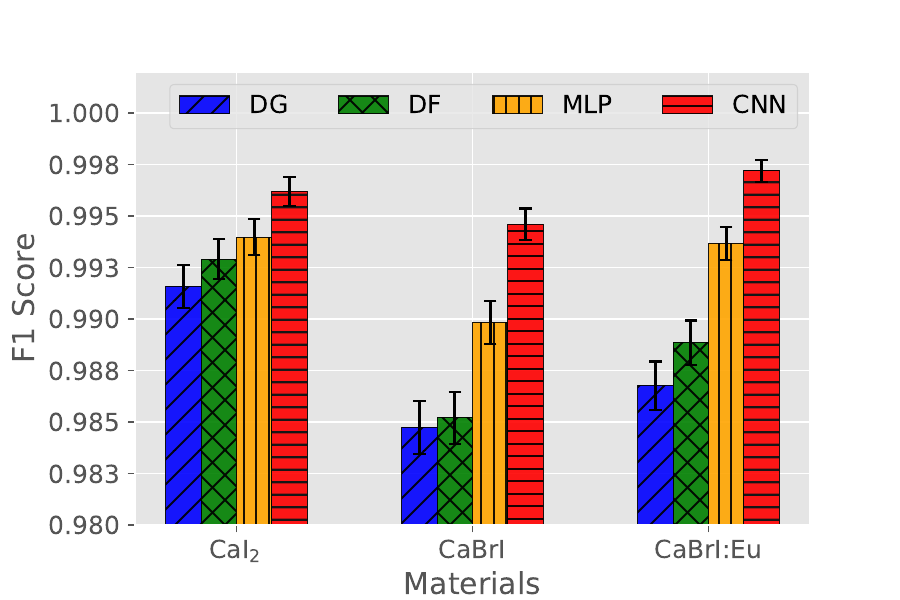}
\caption{PSD performance for the methods.}
\label{fig:mate-f1}
\end{figure}

The confusion matrices of the four analysis methods are summarized in Table \ref{tb:cm}. Using the confusion matrices, the F1 scores were calculated using Eq. \ref{eq:f1}.
The F1 scores, which were used for PSD performance in this study, are shown in Fig. \ref{fig:mate-f1}.
Both of the machine learning analysis methods (MLP and CNN) had better results than the conventional methods (DG and DF).
For all the crystals used in this study, the CNN achieved the best F1 score.
In the comparison of F1 scores for the three crystals, CaI$_2$ had the best F1 score, except in the CNN, and CaBrI:Eu had the best F1 score in the CNN.

\begin{table}[t]
  \centering
  \caption{Confusion matrices of the four analysis methods.}
  \begin{tabular}{ccllll}
    Methods & Materials &   TN &    FP &    FN &    TP  \\ \hline
    DG  &   CaI$_2$        &   2608 &  14 &    50 &    3777  \\
        &   CaBrI       &   2751 &  43 &    98 &    4547  \\
        &   CaBrI:Eu    &   2687 &  27 &    100 &   4736  \\ 
    DF  &   CaI$_2$        &   2614 &  8 &     46 &    3781  \\
        &   CaBrI       &   2741 &  53 &    84 &    4561  \\
        &   CaBrI:Eu    &   2696 &  18 &    89 &    4747  \\ 
    MLP &   CaI$_2$        &   2604 &  18 &    28 &    3799  \\
        &   CaBrI       &   2774 &  20 &    74 &    4571  \\
        &   CaBrI:Eu    &   2698 &  16 &    45 &    4791 \\ 
    CNN &   CaI$_2$        &   2610 &  12 &    17 &    3810  \\
        &   CaBrI       &   2774 &  20 &    30 &    4615 \\
        &   CaBrI:Eu    &   2704 &  10 &    17 &    4819 \\ \hline    
  \end{tabular}
  \label{tb:cm}
\end{table}

\subsubsection{PSD performance for the low energy region}
\begin{figure}[ht]
\centering
\includegraphics[keepaspectratio, width=\linewidth]
{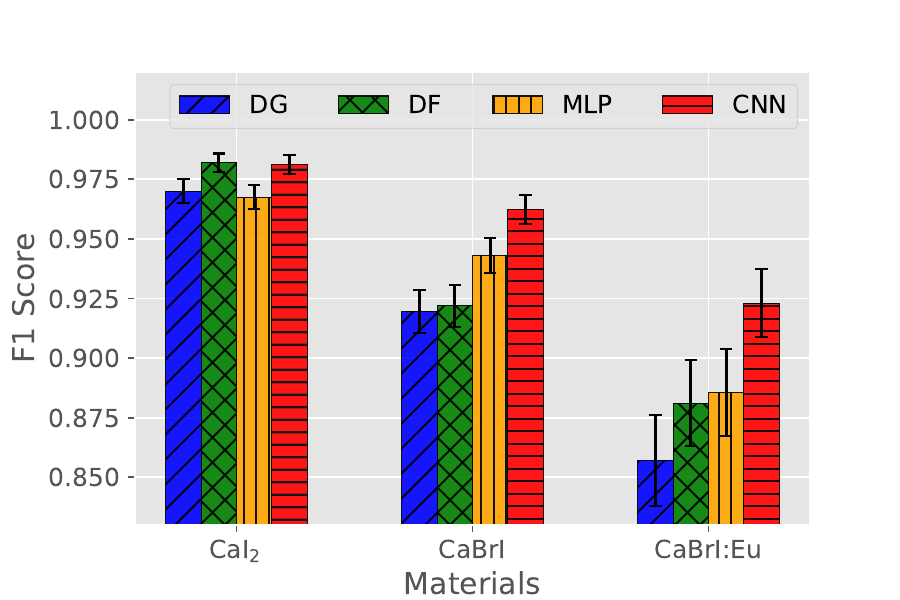}
\caption{PSD performance for lower energy (100 --200~keVee) using the methods.}
\label{fig:mate-f1-le}
\end{figure}

In the low-energy range, PSD is more difficult because very few photoelectrons are produced by the scintillator and the statistical fluctuations have a great effect on the PSD index.
Additionally, the signal is small and the effect of electrical noise is great. In this study, PSD performance was also compared in the low-energy region (100--200~\textcolor{black}{keVee}).
The confusion matrices and F1 score of the four analysis methods are summarized in Table \ref{tb:cm_le} and Fig. \ref{fig:mate-f1-le}. The large error bars in the F1 score for CaBrI:Eu were caused by the small number of $\alpha$-ray events at 100--200 \textcolor{black}{keVee} in CaBrI:Eu compared with the $\alpha$-ray events in the other samples.

Consequently, in the low-energy region, CaI$_2$ had the best F1 score for all analysis methods.
The reason that CaI$_2$ had the highest F1 score in the low-energy region can be attributed to the fact that CaI$_2$ had the highest light output.
Comparing the methods for CaI$_2$, DF and CNN were equally good.
Comparing the methods for CaBrI and CaBrI:Eu, PSD performance was better when the machine learning methods were used, with CNN the best and MLP second best. This trend was the same for the analysis in the entire energy range.

% These results suggest that for scintillators with a high light output such as CaI$_2$, the influence of the analysis method on the PSD accuracy in the low energy region is not significant.
These results suggest that, irrespective of the analysis methods, scintillators with a very high light output, such as CaI$_2$, achieve excellent PSD accuracy in the low-energy range.
By contrast, for scintillators with lower light outputs, such as CaBrI and CaBrI:Eu, PSD accuracy depends on the analysis method, and the machine learning methods achieve better PSD performance than conventional methods.
Although it had the lowest light output (51,000 ph/MeV) in this study, even CaBrI is regarded as a high light output material in the repertoire of conventional scintillation materials. In future work, the machine learning methods developed in this study will be applied to scintillation materials with a low light output, such as CaF$_2$:Eu.

\begin{table}[htbp]
  \centering
  \caption{Confusion matrices of the four analysis methods in the low-energy region (100~keVee--200~keVee).}
  \begin{tabular}{ccrrrr}
    Methods & Materials &   TN &    FP &    FN &    TP  \\ \hline
    DG  &   CaI$_2$     &   1266 &  3 &     33 &    581  \\
        &   CaBrI       &   1248 &  26 &    52 &    446  \\
        &   CaBrI:Eu    &   1221 &  29 &    22 &   153  \\ 
    DF  &   CaI$_2$     &   1262 &  7 &     15 &    599  \\
        &   CaBrI       &   1247 &  27 &    49 &    449  \\
        &   CaBrI:Eu    &   1232 &  18 &    23 &    152  \\ 
    MLP &   CaI$_2$     &   1247 &  22 &    18 &    596  \\
        &   CaBrI       &   1233 &  17 &    41 &    481  \\
        &   CaBrI:Eu    &   1240 &  10 &    28 &    147 \\ 
    CNN &   CaI$_2$     &   1259 &  13 &    10 &    601  \\
        &   CaBrI       &   1264 &  27 &    10 &    471 \\
        &   CaBrI:Eu    &   1243 &  19 &    7 &    156 \\ \hline    
  \end{tabular}
  \label{tb:cm_le}
\end{table}

%% file: 5_conclusion.tex
\section{Conclusion}
\label{sec:conc}
In this study, PSD performance was compared for two conventional methods and two machine learning methods. A dual gate method and DF method were applied as conventional methods, and an MLP %network
and CNN as machine learning methods.
The scintillator crystals used in this study were CaI$_2$, CaBrI, and Eu 2~\% doped CaBrI, grown using the vertical Bridgman--Stockbarger method. 
The Bridgman-grown CaI$_2$, CaBrI, and CaBrI:Eu crystals had the light output of 107,000, 51,000, and 79,000 ph/MeV, respectively. 
The PSD results over the entire energy range demonstrated that the machine learning methods achieved better F1 scores than the conventional methods for all crystals; CNN was the best and the MLP %network
second best. 
CaBrI:Eu had the best F1 score when the CNN was applied.
For the lower energy region of 100--200~\textcolor{black}{keVee}, CaI$_2$ was the best for all analysis methods because it had the highest light output. The CNN and DF method were the best for CaI$_2$ in the range 100--200~\textcolor{black}{keVee}.
The results for CaBrI and CaBrI:Eu in this energy range showed the same trend as in the full energy range.
Our results showed that for scintillators with a very high light output, such as CaI$_2$, the PSD accuracy was excellent in the low-energy region, independent of the analysis method.
For scintillators with a lower light output, such as CaBrI and CaBrI:Eu, the machine learning methods achieved better PSD accuracy than the conventional method.
Future work includes the application of machine learning methods to analyze the PSD of low light output scintillators, such as CaF$_2$:Eu.

%% file: 0_main.bbl
\begin{thebibliography}{10}
\expandafter\ifx\csname url\endcsname\relax
  \def\url#1{\texttt{#1}}\fi
\expandafter\ifx\csname urlprefix\endcsname\relax\def\urlprefix{URL }\fi
\expandafter\ifx\csname href\endcsname\relax
  \def\href#1#2{#2} \def\path#1{#1}\fi

\bibitem{gas}
V.~Chepel, H.~Ara{\'{u}}jo,
  \href{https://doi.org/10.1088/1748-0221/8/04/r04001}{Liquid noble gas
  detectors for low energy particle physics}, Journal of Instrumentation 8~(04)
  (2013) R04001--R04001.
\newblock \href {http://dx.doi.org/10.1088/1748-0221/8/04/r04001}
  {\path{doi:10.1088/1748-0221/8/04/r04001}}.
\newline\urlprefix\url{https://doi.org/10.1088/1748-0221/8/04/r04001}

\bibitem{bolometer}
A.~Alessandrello, V.~Bashkirov, C.~Brofferio, C.~Bucci, D.~Camin, O.~Cremonesi,
  E.~Fiorini, G.~Gervasio, A.~Giuliani, A.~Nucciotti, M.~Pavan, G.~Pessina,
  E.~Previtali, L.~Zanotti,
  \href{https://www.sciencedirect.com/science/article/pii/S037026939701544X}{A
  scintillating bolometer for experiments on double beta decay}, Physics
  Letters B 420~(1) (1998) 109--113.
\newblock \href
  {http://dx.doi.org/https://doi.org/10.1016/S0370-2693(97)01544-X}
  {\path{doi:https://doi.org/10.1016/S0370-2693(97)01544-X}}.
\newline\urlprefix\url{https://www.sciencedirect.com/science/article/pii/S037026939701544X}

\bibitem{CaI2}
K.~Kamada, T.~Iida, T.~Ohata, M.~Yoshino, S.~Hayasaka, Y.~Shoji, S.~Sakuragi,
  S.~Kurosawa, Y.~Yokota, Y.~Ohashi, A.~Yoshikawa,
  \href{https://www.sciencedirect.com/science/article/pii/S0272884217310714}{Single
  crystal growth and scintillation properties of ca(cl, br, i)2 single
  crystal}, Ceramics International 43 (2017) S423--S427, the 10th Asian Meeting
  on Electroceramics (AMEC-10).
\newblock \href
  {http://dx.doi.org/https://doi.org/10.1016/j.ceramint.2017.05.249}
  {\path{doi:https://doi.org/10.1016/j.ceramint.2017.05.249}}.
\newline\urlprefix\url{https://www.sciencedirect.com/science/article/pii/S0272884217310714}

\bibitem{CaI-PSD}
T.~Iida, K.~Kamada, M.~Yoshino, K.~J. Kim, K.~Ichimura, A.~Yoshikawa,
  \href{https://www.sciencedirect.com/science/article/pii/S0168900219311271}{High-light-yield
  calcium iodide (cai2) scintillator for astroparticle physics}, Nuclear
  Instruments and Methods in Physics Research Section A: Accelerators,
  Spectrometers, Detectors and Associated Equipment 958 (2020) 162629,
  proceedings of the Vienna Conference on Instrumentation 2019.
\newblock \href {http://dx.doi.org/https://doi.org/10.1016/j.nima.2019.162629}
  {\path{doi:https://doi.org/10.1016/j.nima.2019.162629}}.
\newline\urlprefix\url{https://www.sciencedirect.com/science/article/pii/S0168900219311271}

\bibitem{CaBrI}
K.~Kamada, T.~Iida, Y.~Furuya, K.~J. Kim, M.~Yoshino, R.~Murakami, Y.~Shoji,
  V.~V. Kochurikhin, A.~Yamaj, S.~Kurosawa, Y.~Ohashi, Y.~Yokota, A.~Yoshikawa,
  \href{https://www.sciencedirect.com/science/article/pii/S1350448718307157}{Crystal
  growth and scintillation properties of eu-doped ca(brxi1-x)2 crystals},
  Radiation Measurements 127 (2019) 106139.
\newblock \href
  {http://dx.doi.org/https://doi.org/10.1016/j.radmeas.2019.106139}
  {\path{doi:https://doi.org/10.1016/j.radmeas.2019.106139}}.
\newline\urlprefix\url{https://www.sciencedirect.com/science/article/pii/S1350448718307157}

\bibitem{CaClBrI}
K.~Kamada, T.~Iida, T.~Ohata, M.~Yoshino, S.~Hayasaka, Y.~Shoji, S.~Sakuragi,
  S.~Kurosawa, Y.~Yokota, Y.~Ohashi, A.~Yoshikawa,
  \href{https://www.sciencedirect.com/science/article/pii/S0272884217310714}{Single
  crystal growth and scintillation properties of ca(cl, br, i)2 single
  crystal}, Ceramics International 43 (2017) S423--S427.
\newblock \href
  {http://dx.doi.org/https://doi.org/10.1016/j.ceramint.2017.05.249}
  {\path{doi:https://doi.org/10.1016/j.ceramint.2017.05.249}}.
\newline\urlprefix\url{https://www.sciencedirect.com/science/article/pii/S0272884217310714}

\bibitem{WaveCatcher}
J.~M. D.~Breton, E.~Delagnes, P.~Rusquart,
  \href{https://ieeexplore.ieee.org/document/7097545}{The wavecatcher family of
  sca-based 12-bit 3.2-gs/s fast digitizers}, 2014 19th IEEE-NPSS Real Time
  Conference (2014) 1--8\href
  {http://dx.doi.org/https://doi.org/10.1109/RTC.2014.7097545}
  {\path{doi:https://doi.org/10.1109/RTC.2014.7097545}}.
\newline\urlprefix\url{https://ieeexplore.ieee.org/document/7097545}

\bibitem{TAMAGAWA2015192}
Y.~Tamagawa, Y.~Inukai, I.~Ogawa, M.~Kobayashi,
  \href{https://www.sciencedirect.com/science/article/pii/S0168900215007111}{Alpha–gamma
  pulse-shape discrimination in gd3al2ga3o12 (gagg):ce3+ crystal scintillator
  using shape indicator}, Nuclear Instruments and Methods in Physics Research
  Section A: Accelerators, Spectrometers, Detectors and Associated Equipment
  795 (2015) 192--195.
\newblock \href {http://dx.doi.org/https://doi.org/10.1016/j.nima.2015.05.052}
  {\path{doi:https://doi.org/10.1016/j.nima.2015.05.052}}.
\newline\urlprefix\url{https://www.sciencedirect.com/science/article/pii/S0168900215007111}

\bibitem{Mizukoshi_2019}
K.~Mizukoshi, T.~Iida, I.~Ogawa, K.~Shimizu, S.~Kurosawa, K.~Kamada,
  M.~Yoshino, A.~Yoshikawa,
  \href{https://doi.org/10.1088/1748-0221/14/06/p06037}{Pulse-shape
  discrimination potential of new scintillator material: La-{GPS}:ce}, Journal
  of Instrumentation 14~(06) (2019) P06037--P06037.
\newblock \href {http://dx.doi.org/10.1088/1748-0221/14/06/p06037}
  {\path{doi:10.1088/1748-0221/14/06/p06037}}.
\newline\urlprefix\url{https://doi.org/10.1088/1748-0221/14/06/p06037}

\bibitem{dpp_psd}
C.~S.p.A., \href{https://www.caen.it/products/dpp-psd}{Digital pulse processing
  for charge integration and pulse shape discrimination} (2021).
\newline\urlprefix\url{https://www.caen.it/products/dpp-psd}

\bibitem{bib:Miyazaki:ResNet}
K.~He, X.~Zhang, S.~Ren, J.~Sun, Deep residual learning for image recognition,
  in: 2016 IEEE Conference on Computer Vision and Pattern Recognition (CVPR),
  2016, pp. 770--778.
\newblock \href {http://dx.doi.org/10.1109/CVPR.2016.90}
  {\path{doi:10.1109/CVPR.2016.90}}.

\bibitem{bib:Miyazaki:batch_norm}
S.~Ioffe, C.~Szegedy, Batch normalization: Accelerating deep network training
  by reducing internal covariate shift, in: Proceedings of the 32nd
  International Conference on International Conference on Machine Learning -
  Volume 37, ICML'15, JMLR.org, 2015, p. 448–456.

\bibitem{birks1951}
J.~B. Birks, {Scintillations from Organic Crystals: Specific Fluorescence and
  Relative Response to Different Radiations}, Proceedings of the Physical
  Society. Section A 64~(10) (1951) 874--877.
\newblock \href {http://dx.doi.org/10.1088/0370-1298/64/10/303}
  {\path{doi:10.1088/0370-1298/64/10/303}}.

\end{thebibliography}
